\documentclass{elsart}

\begin{document}

\begin{frontmatter}

\title{What is the value of an observable between pre- and postselection?}
\author{Lars M. Johansen}

\address{Department of Technology,
Buskerud University College, N-3601 Kongsberg, Norway}
\date{\date}
\ead{lars.m.johansen@hibu.no}

\maketitle

\begin{abstract}

Hall's recent derivation of an exact uncertainty relation [Phys.
Rev. A64, 052103 (2001)] is revisited. It is found that the Bayes
estimator of an observable between pre- and postselection equals
the real part of the weak value. The quadratic loss function
equals the expectation of the squared imaginary part of the weak
value.

\end{abstract}


\end{frontmatter}

Hall recently solved the problem of finding the most efficient
estimator for an observable on basis of a measurement of another,
incompatible observable \cite{Hall-Exacuncerela:01}. As a
consequence, he also found an ``exact uncertainty relation". In
this short note, we revisit Hall's derivation and provide a new
interpretation of the results. In particular, we demonstrate that
the weak value of an observable is the most efficient estimator of
an observable between preselection and postselection. This result
is consistent with the fact that weak values are measured in weak
measurements where the interaction between the system and the
measurement apparatus is weak
\cite{Aharonov+AlbertETAL-ResuMeasCompSpin:88}. Also, we show that
the loss function, which expresses the uncertainty in the
estimate, equals the squared imaginary part of the weak value.
This result is closely related to the exact uncertainty relation
\cite{Hall-Exacuncerela:01}.

It was demonstrated by Aharonov \emph{et al.}\
\cite{Aharonov+AlbertETAL-ResuMeasCompSpin:88} that if a ``weak
measurement" is performed of an observable $\hat{a}$ between
preselection of a state $\mid \psi \rangle$ and postselection of a
state $\mid b \rangle$, then what is measured is the `` weak
value" of the observable,
\begin{equation}
    \label{eq:pureweakvalue}
    a_w(b) = {\langle b \mid \hat{a} \mid \psi \rangle \over \langle
    b \mid \psi \rangle}.
\end{equation}
More precisely, the expectation value of the meter reading equals
the real part of the weak value. It has also been shown that the
imaginary part of the weak value can be observed in a weak
measurement \cite{Aharonov+Vaidman-Propquansystduri:90}. $a_w$ has
also been investigated in other settings. Thus, the real part of
$a_w$ has been interpreted as the ``local value" of the observable
$\hat{a}$
\cite{Wan+Sumner-Quanpotespatdist:88,%
Holland-QuanTheoMoti:93,%
Cohen-Locavaluquanmech:96,%
Muga+PalaoETAL-Averlocavaluloca:98}.

In recent years, weak values and weak measurements have been
applied to a variety of problems in quantum mechanics. For
example, there has been a long discussion on what is the correct
way of calculating the time spent by a particle in a tunneling
region. Steinberg \cite{Steinberg-MuchTimeDoesTunn:95} has pointed
out that in the tunneling time problem we are actually dealing
with a pre- and postselected system. Steinberg also proposed to
use weak values in determining the tunneling time. Weak values
have been applied in the interpretation of the double slit
experiment \cite{Wiseman-Direobsemometran:03}. It has been shown
that weak values apply to fiber optical networks
\cite{Brunner+AcinETAL-Optitelenetwweak:03}.  Weak values have
also been observed in quantum optical experiments
\cite{Ritchie+StoryETAL-RealMeasWeakValu:91,Parks+CullinETAL-ObsemeasoptiAhar:98}.
Recently, it was proposed to test Hardy's paradox
\cite{Hardy-QuanMechLocaReal:92} experimentally through weak
measurements \cite{Aharonov+BoteroETAL-ReviHardpara:02}. This
suggestion has aroused a lot of interest
\cite{Scientist-Curicuri:03}.

Following Hall \cite{Hall-Exacuncerela:01}, we use the strategy of
finding the best possible \emph{guess} of $\hat{a}$, i.e., the
best possible estimator. In quantum mechanics, observables are
stochastic quantities. In standard parametric estimation theory,
the aim is usually to estimate a fixed, but unknown parameter. It
is therefore not very well suited to solving our problem. On the
other hand, Bayesian estimation theory is designed for the
estimation of stochastic parameters. It can also take into account
prior information about the parameter. We shall consider the
preselected state as prior information about the observable. We
consider in general a mixed state $\hat{\rho}$ as prior.

The most efficient estimator, or the Bayes estimator, is defined
as the estimator that minimizes a given loss function. The loss
gives a measure of the deviation between the estimator and the
intrinsic observable to be estimated. There are several different
loss functions in use in Bayesian estimation theory. The quadratic
loss is most frequently used, and can be compared with the
prevalence of variance as the most frequently used measure of
spread around an expectation value.

We introduce $\hat{\theta}$ as an estimator for $\hat{a}$. As
usual in estimation theory, this estimator must be a function of
the measurement, i.e., the postselection operator $\hat{b}$.
Therefore, these operators commute, $[ \, \hat{\theta},\hat{b} \,]
= 0$.

The quadratic loss is
\begin{equation}
    \label{eq:loss}
    L(\hat{\theta}) = \langle (\hat{\theta}-\hat{a})^2 \rangle,
\end{equation}
where $\langle \hat{O} \rangle = \mathrm{Tr} \hat{\rho} \,
\hat{O}$. Next, we introduce the following operator compatible
with $\hat{b}$,
\begin{equation}
    \label{eq:weakvalueop}
    \hat{\alpha} = \int db \; \alpha(b) \mid b \rangle
    \langle b \mid,
\end{equation}
where
\begin{equation}
    \label{eq:weakvalue}
    \alpha(b) = {\langle b \mid \hat{a} \hat{\rho} \mid b \rangle
    \over \langle b \mid \hat{\rho} \mid b \rangle}.
\end{equation}
We write $\hat{\alpha} = \hat{\mu} + i \, \hat{\sigma}$, where
\begin{eqnarray}
    \label{eq:bayes}
    \hat{\mu} &=& {1 \over 2} (\hat{\alpha} + \hat{\alpha}^\dag) =
    \int db \; \mu(b) \mid b \rangle \langle b \mid,
    \\
    \hat{\sigma} &=& {1 \over 2i} \, (\hat{\alpha} - \hat{\alpha}^\dag)
    = \int db \; \sigma(b) \mid b \rangle \langle b \mid.
\end{eqnarray}
It follows that
\begin{eqnarray}
    \mu(b) &=& {1 \over 2} [\alpha(b) + \alpha^*(b)],\\
    \sigma(b) &=& {1 \over 2i} \, [\alpha(b) - \alpha^*(b)].
\end{eqnarray}
The loss can be written in the form \cite{Hall-Exacuncerela:01}
\begin{equation}
    \label{eq:halloss}
    L(\hat{\theta}) = \langle \hat{a}^2 \rangle - \langle
    \hat{\theta}^2 \rangle + \langle (\hat{\theta} -
    \hat{\mu})^2 \rangle.
\end{equation}
It is minimized for the estimator $\hat{\theta} = \hat{\mu}$. This
is the Bayes estimator, and it can be shown to be unbiased,
$\langle \hat{\mu} \rangle = \langle \hat{a} \rangle$.

Hall has proposed to interpret $\hat{\mu}$ as the ``classical
component" of the observable $\hat{a}$ \cite{Hall-Exacuncerela:01}
(see also
\cite{Luo-StatLocaValuQuan:02,Luis-Phasdistclascomp:03}). However,
this terminology is not very appropriate. If we restrict the
attention to pure states $\hat{\rho} = \mid \psi \rangle \langle
\psi \mid$, then we see that $\mu(b) = \mathrm{Re} \, a_w(b)$. In
other words, the Bayes estimator equals the real part of the weak
value. Weak values are notoriously known for their bizarre, and
sometimes nonclassical properties. For example, the weak value of
kinetic energy may be negative
\cite{Aharonov+PopescuETAL-MeasErroNegaKine:93}. It is therefore
hardly justifiable to interpret the Bayes estimator as the
``classical component" of the observable.

In Ref. \cite{Cohen-Locavaluquanmech:96} $\mu(b)$ was interpreted
as the ``local value" of the observable $\hat{a}$, and
$\sigma^2(b)$ was interpreted as the ``local variance". The latter
claim was criticized in Ref.
\cite{Muga+PalaoETAL-Averlocavaluloca:98}.

The loss gives a measure of how well the estimator approaches the
intrinsic observable $\hat{a}$. We now find a lower bound on the
loss. We follow Hall \cite{Hall-Exacuncerela:01} and introduce the
vectors $\mid \mu \rangle = \hat{\rho}^{1/2} \hat{a} \mid b
\rangle$ and $\mid \nu \rangle = \hat{\rho}^{1/2} \mid b \rangle$.
By using the Schwarz inequality $\mid \langle \mu \mid \nu \rangle
\mid^2 \le \langle \mu \mid \mu \rangle \langle \nu \mid \nu
\rangle$, we find that
\begin{equation}
    \label{eq:schwarz}
    \langle \hat{\alpha}^2 \rangle = \langle \hat{\mu}^2 \rangle +
    \langle \hat{\sigma}^2 \rangle\le \langle \hat{a}^2 \rangle.
\end{equation}
Equality is reached for pure states \cite{Hall-Exacuncerela:01}.
From Eq.\ (\ref{eq:halloss}) follows the inequality
\begin{equation}
    \label{eq:cbound}
    L(\hat{\theta}) \ge \langle \hat{a}^2 \rangle - \langle
    \hat{\mu}^2 \rangle.
\end{equation}
The lower bound is obtained for the Bayes estimator $\hat{\mu}$.
Combining Eqs.\ (\ref{eq:schwarz}) and (\ref{eq:cbound}) yields
the inequality
\begin{equation}
L (\hat{\theta}) \ge \langle \hat{\sigma}^2 \rangle.
\end{equation}
The loss is bounded from below by the squared imaginary part of
the weak value. The lower bound is attained by the Bayes estimator
on pure states. The equality which is attained for the Bayes
estimator can be rewritten as an exact uncertainty relation in the
case when $\hat{a}=\hat{p}$ and $\hat{b}=\hat{q}$
\cite{Hall-Exacuncerela:01}. So, the exact uncertainty relation
can be interpreted as saying that the weak value is the Bayes
estimator on a pre- and postselected ensemble.

To answer the question posed in the title, we see that a quantum
observable has no definite value between pre- and postselection.
Even the most efficient estimator has a nonvanishing loss
function. However, in the special case when $\langle
\hat{\sigma}^2 \rangle$ vanishes, there exists an exact estimate.
This requires that $\sigma(b)$, the imaginary part of the weak
value, must vanish.

In conclusion, we have shown that the Bayes estimator on a pre-
and postselected ensemble is equal to the real part of the weak
value. This result agrees nicely with the fact that weak values
are observed in weak measurements on pre- and postselected
ensembles. We found that a lower bound on the quadratic loss is
given by the expectation of the squared imaginary part of the weak
value. The bound is reached for the Bayes estimator on pure
states.

\section*{Acknowledgments}

This work was supported by a grant from Buskerud University
College.


\begin{thebibliography}{10}
\expandafter\ifx\csname url\endcsname\relax
  \def\url#1{\texttt{#1}}\fi
\expandafter\ifx\csname
urlprefix\endcsname\relax\def\urlprefix{URL }\fi

\bibitem{Hall-Exacuncerela:01}
M.~J.~W. Hall, Exact uncertainty relations, Phys. Rev. A 64 (2001)
  052103--1--052103--10.

\bibitem{Aharonov+AlbertETAL-ResuMeasCompSpin:88}
Y.~Aharonov, D.~Z. Albert, L.~Vaidman, How the result of a
measurement of a
  component of the spin of a spin-$1 \over 2$ particle can turn out to be 100,
  Phys. Rev. Lett. 60~(14) (1988) 1351--1354.

\bibitem{Aharonov+Vaidman-Propquansystduri:90}
Y.~Aharonov, L.~Vaidman, Properties of a quantum system during the
time
  interval between two measurements, Phys. Rev. A 41~(1) (1990) 11.

\bibitem{Wan+Sumner-Quanpotespatdist:88}
K.~K. Wan, P.~J. Sumner, Quantum potential and the spatial
distribution of
  observable values, Phys. Lett. A 128~(9) (1988) 458--462.

\bibitem{Holland-QuanTheoMoti:93}
P.~R. Holland, The Quantum Theory of Motion, Cambridge University
Press,
  Cambridge, 1993.

\bibitem{Cohen-Locavaluquanmech:96}
L.~Cohen, Local values in quantum mechanics, Phys. Lett. A 212
(1996) 315--319.

\bibitem{Muga+PalaoETAL-Averlocavaluloca:98}
J.~G. Muga, J.~P. Palao, R.~Sala, Average local values and local
variances in
  quantum mechanics, Phys. Lett. A 238~(2-3) (1998) 90--94.

\bibitem{Steinberg-MuchTimeDoesTunn:95}
A.~M. Steinberg, How much time does a tunneling particle spend in
the barrier
  region, Phys. Rev. Lett. 74~(13) (1995) 2405--2409.

\bibitem{Wiseman-Direobsemometran:03}
H.~M. Wiseman, Directly observing momentum transfer in twin-slit
"which-way"
  experiments, Phys. Lett. A 311 (2003) 285--291.

\bibitem{Brunner+AcinETAL-Optitelenetwweak:03}
N.~Brunner, A.~Acin, D.~Collins, N.~Gisin, V.~Scarani, Optical
telecom networks
  as weak quantum measurements with post-selection, quant-ph/0306108 .

\bibitem{Ritchie+StoryETAL-RealMeasWeakValu:91}
N.~W.~M. Ritchie, J.~G. Story, R.~G. Hulet, Realization of a
measurement of a
  "weak value", Phys. Rev. Lett. 66~(9) (1991) 1107--1110.

\bibitem{Parks+CullinETAL-ObsemeasoptiAhar:98}
A.~D. Parks, D.~W. Cullin, D.~C. Stoudt, Observation and
measurement of an
  optical aharonov-albert-vaidman effect, Proc. R. Soc. Lond. A 454 (1998)
  2997--3008.

\bibitem{Hardy-QuanMechLocaReal:92}
L.~Hardy, Quantum mechanics, local realistic theories, and
lorentz-invariant
  theories, Phys. Rev. Lett. 68~(20) (1992) 2981--2984.

\bibitem{Aharonov+BoteroETAL-ReviHardpara:02}
Y.~Aharonov, A.~Botero, S.~Popescu, B.~Reznik, J.~Tollaksen,
Revisiting hardy's
  paradox: counterfactual statements, real measurements, entanglement and weak
  values, Phys. Lett. A 301 (2002) 130--138.

\bibitem{Scientist-Curicuri:03}
Curioser and curioser, New Scientist 178~(2394) (2003) 28.

\bibitem{Luo-StatLocaValuQuan:02}
S.~Luo, Statistics of local value in quantum mechanics, Int. J.
Theor. Phys.
  41~(9) (2002) 1713--1730.

\bibitem{Luis-Phasdistclascomp:03}
A.~Luis, Phase-space distributions and the classical component of
quantum
  observables, Phys. Rev. A 67 (2003) 064101.

\bibitem{Aharonov+PopescuETAL-MeasErroNegaKine:93}
Y.~Aharonov, S.~Popescu, D.~Rohrlich, L.~Vaidman, Measurements,
error, and
  negative kinetic energy, Phys. Rev. A 48~(6) (1993) 4084--90.

\end{thebibliography}
\end{document}